# IMPACT GENERATED SHOCKWAVES ARE PROPOSED FOR THE ORIGIN OF SUNSPOTS TO EXPLAIN THE DETECTED PLANETARY EFFECTS ON THE SOLAR ACTIVITY

Jozsef Garai

ABSTRACT

Five new correlations between sunspot activity and orbiting position of the Jovian planets are detected. In order to explain these correlations it is suggested that the resonance of the outer planets destabilizes the orbit of Kuiper Belt Objects and generates a cyclical impact frequency on the Sun. The vaporization of the object initiates a shock way disrupting the upwelling of the plasma resulting in a sunspot formation. The proposed model is able to explain the length of the cycle, the latitude distribution of the sunspots and the extremely long term stability of the cycles. Calculating the positions of the Jovian planets at conjunction and opposition allows the long term prediction of the solar activity.

## 1. INTRODUCTION

The close coincidence of the sideral period of the largest planet, Jupiter (11.87), with the about eleven-year cylicity of the solar activity has been known for long time. In the middle nineteen-century R. Wolf and R.C. Carrington suggested (Smythe & Addy, 1977) that the gravitational pull of the planets might trigger the solar activity by the tides raised on the Sun. Tidal model of Venus, Earth and Jupiter was suggested to explain solar activity based on the correlation which was detected between the calculated tidal height fluctuation of the three planets and sunspot numbers (Wood & Wood, 1965; Wood, 1972). This model was entirely discredited because did not provide a credible physical model to explain solar magnetism, Maunder minimum (Smythe & Eddy, 1977) and the butterfly distribution of the spots. Even the detected 11.08 y. peridoicity of the tidal planets were questioned and it was suggested that the correlation is an artifact of the calculations (Okal & Anderson, 1975).

The known and accepted periodicities in the solar cycle determined form the entire record (1749-1990) are $11 \pm 1.5$ $y$ and $10 \pm 0.1$ $y$ (Cohen & Lintz, 1974; Berger et al. 1990; Donahue & Baliunas, 1992). Other periodicities, like the 8.3 y, and the 95.8 y with less significance had been detected but the existence of these cycles are still debated (Wallenhorst, 1982). The planetary orbital periods coincides with the above cycles are V-E$_{(resonance)}$ = 8.1 y, J/S = 9.929 y, and U/N = 85.698 y (e.g. Moerth & Schlamminger, 1978). There is a small but consistent correlation between the Zurich relative daily sunspot numbers for the years 1850-1960 (Waldmeier, 1961) and the sideral period of the planet Mercury (Bigg, 1967). Despite the numerous correlations between planetary orbiting periods and solar activity current literature assumes that planets do not affect solar activity and internal conditions are responsible for the solar cycle.

Some features of the solar activity is not consistent with the internal origin of the sunspots. The cyclical appearance and well defined pattern of the sunspots is nonrandom. The convection of the hot plasma which rises up from the interior sun and spreads out across the surface and then cools and sinks inward is without any doubt random. Random process can not generate a non-random process. Thus the sunspot formation should not originate from and relate to random plasma convection.

The 11 and 22 years cycles in the solar activity had been detected for hundreds of millions of years (Williams, 1981). Beside astronomical driven cycle, there is no known other process able to sustain this long-term stability. The only astronomical cycles consistent with length of the solar cycle are planetary orbiting periods. The correlation between solar activity and planetary orbits is revisited and investigated.

## 2. CORRELATIONS BETWEEN SOLAR ACTIVITY AND THE POSITIONS OF THE JOVIAN PLANETS

The only long range effect of the planets which might have influence on solar activity is gravity. The orbiting of the planets causes fluctuations in the gravity field of the solar system, which is dominated by Jupiter and Saturn. The field is the most intense when the gravitation of the two planets is superimposed at conjunction and opposition. Calculating the heliocentric longitudes of the Jovian planets (Smith, 1981) at heliospheric conjunctions and oppositions of Jupiter and Saturn (HCOJS) correlations between solar activity and planetary positions are investigated.

The Greenwich data set of the yearly average latitude of the sunspots shows systematic behavior in relations to HCOJS (Fig. 1). The lowest latitudes do coincide with the time of HCOJS in seven cycles (15-20). This pattern is not present in the first three solar cycles (12-14) of the data. In the period of 1700-2008 thirty HCOJS occurred and 28 solar cycles -4 through 23 was observed. Solar cycles 5 and 12 covered two HCOJS. The three "irregular" cycles in which the lowest latitudes did not coincide with the HCOJS follows a cycle which covered two HCOJS. It is assumed that if an HCOJS is skipped then irregularities in the following three cycles could

occur. Based on this assumption solar cycles following the HCOJS of 1703; 1713; 1802; 1812; 1822; 1881; 1892; 1902 can be identified as irregular. I will call the rest of the cycles regular. The assignments of HCOJS to solar cycles are shown in Table 1. The investigated data set contains 20 regular and 8 irregular cycles. The effect of HCOJS on these two groups of cycles is investigated separately.

Representing the position of Uranus and Neptune by the angle between Jupiter and the planet at the time of HCOJS and investigating the affect of these orbiting components on the maximum yearly average sunspot number indicates the presence of the following correlations:

- small heliocentric longitudes of Jupiter at the time of HCOJS correlate to low solar maximums and high latitudes to high activity
- small angles between Jupiter and Uranus at the time of the HCOJS correlate to high solar activities while the big angles correlate to low activities
- small angles between Jupiter and Neptune at the time of the HCOJS correlate to low activities while the big angles correlate to high activities

The correlations are:
$$MSN_{(R)} = 97.88 + 0.1255 \times JHL \qquad R = 0.42 \qquad (1)$$
where MSN is the Maximum Yearly Average Sunspot Number, JHL is the Heliocentric Longitude of Jupiter at the time of HCOJS and R is the correlation coefficient. Subscript R refers to regular cycles.
$$MSN_{(R)} = 143.05 - 0.2196 \times JU \qquad R = 0.41 \qquad (2)$$
where, JU is the angle between Jupiter and Uranus at the time of HCOJS.
$$MSN_{(R)} = 106.94 + 0.1835 \times JN \qquad R = 0.33 \qquad (3)$$
where JN is the angle between Jupiter and Neptune at the time of HCOJS. The correlations are shown on Fig. 2.

Combining the three planetary orbiting effects gives the Maximum Sunspot Number as:
$$MSN_{(R)} = 97.62 + 0.1545 \times JHL - 0.2783 \times JU + 0.2144 \times JN. \qquad (4)$$
The correlation coefficient is 0.747 and the standard error is 21.05. For irregular cycles:
$$MSN_{(I)} = 52.78 + 0.0462 \times JHL + 0.0773 \times JU - 0.1010 \times JN. \qquad (5)$$
The correlation coefficient is 0.585 and standard error is 13.48.

Equations (4) and (5) have upper (192) and lover (36) limit on the Maximum Yearly Average Sunspot Number. Thus the determined correlations do not applicable to the period of the Maunder minimum indicating that additional term/s might be required for longer and more precise description of the solar activity.

The length of the cycle (minima) (Rogers et al., 2006) correlates to the same variables as:

$$L_{(R)} = 9.946 + 9.67 \times 10^{-4} \times JHL + 9.20 \times 10^{-3} \times JU - 2.26 \times 10^{-3} \times JN, \qquad (6)$$

where L is the length of the cycle in years. The correlation coefficient is 0.48.

Using equations (4) and (5) the maximum sunspot numbers are calculated from the orbiting parameters of the Jovian planets. The calculated values of cycles -4 through 26 are plotted against observations (Fig. 3). The agreement is reasonably good.

## 3. INTERPRETATION OF THE CORRELATIONS

The presented correlations between solar activity and the position of the Jovian planets eliminate the likelihood that these correlations are random coincidences. No known internal solar mechanism can explain how Uranus and Neptune could have any direct effect on the Sun. However the observed correlations can be explained by impact generated sunspot formation (Garai, 2001).

It is suggested that the 9.9 y cycle of HCOJS destabilize the orbits of solar system bodies resulting in an impact flux on the Sun with the same cycle. The relative position of Uranus and Neptune to Jupiter slightly modifies this primarily cycle. The conjunction of Jupiter and Neptune at the time of HCOJS results in lower activity and longer cycle. On the other hand, the conjunction of Uranus and Jupiter at the time of HCOJS increases the intensity of the cycle and maintains the fundamental 9.9 y periodicity of the cycle. The solar activity following the triple conjunctions of J-S-U in 1762, and again in 1940, was very high in the following two cycles.

If the heliospheric longitude of Neptune is within $10^0$ of Jupiter's heliospheric longitude while the activity is low, as occurred in 1703, 1792 and 1881, then the resulting depression and delay in the solar activity can cause the following two cycles to fully overlap and therefore not be noticed as independent cycle. This phenomenon did not occur in 1971 because the solar activity was too high at that time. These overlapping cycles modify the fundamental 9.9 y length of the solar cycle to the observed 11.03 y.

It is also suggested that the objects hitting the Sun originate from the Kuiper Belt. The inclination of the Kuiper Belt Objects is between 0-35º (Luu and Jewitt, 2002; Jewitt et al., 2008) which is consistent with the observed latitude distribution of the sunspots.

The flux of the projectiles moving towards the Sun is further affected by the gravitational attraction of the inner planets, which explains why the orbiting periods of the inner planets are also detectable in the solar activity.

The influence of the solar cycle was so much stronger 680 Myr ago than today (Williams,

1981). The weakening of the cycle can be explained by the depletion of the KBO. The well established relationship between mean star activity, age and rotational period (Ducan et al., 1991; Henry et al., 1996) is also consistent with a depletion and impact accretion process.

An impact model is consistent with the cooler temperature of the sunspots and explains why most of the missing radiated energy at the site of the sunspots has never been detected (Rast et al., 1999).

The two components cooling of the sunspots is also consistent with the observed temperature differences in the sunspot umbra which can not be explained by a continuum model (Van Ballegooijen, 1984).

The observed butterfly distribution of the sunspots can be shown consistent with the proposed impact model. The impact free zone around the solar equator at the beginning of the cycle is the result of the shielding effect of Jupiter and Saturn.

## 4. CONCLUSIONS

Five new correlations between the orbiting position of the Jovian planets and solar activity is reported. These additional correlations between planetary positions and sunspot maximum are convincingly demonstrates that the planetary effects on solar activity are real and not an artifact.

Impact generated shock wave is proposed for the formation of sunspots to explain the affect of the Jovian planets on solar activity. It is also suggested that the objects are originates from the Kuiper Belt. This model is able to explain the length of the cycle, the latitude distribution of the sunspots, and the long term stability of the cycles. The impact mechanism is also consistent with the non-random nature of the sunspot distribution, with the missing radiated energy, with the two components cooling of the sunspots and with the relationship between mean star activity, age and rotational period.

The detected correlations allow calculating the maximum sunspot number from planetary positions. The calculated values reproduce the observed sunspot's activity reasonably well for the investigated 300 years. The presented correlation is the first one which is able to give reasonable prediction for longer term solar activity.

I would like to thank to Stanley Adler for reading the manuscript and making comments.

TABLE. 1 The time of the heliospheric conjunction and opposition of Jupiter and Saturn and the corresponding solar cycles.

| Time of HCOJS | Year of the Maximum | Solar Cycle | Characteristic |
|---|---|---|---|
| 1702.8 | 1705.5 | -4 | I |
| 1713.2 | 1718.2 | -3 | I |
| 1723.0 | 1727.5 | -2 | R |
| 1732.6 | 1738.7 | -1 | R |
| 1742.7 | 1750.3 | 0 | R |
| 1752.5 | 1761.5 | 1 | R |
| 1762.4 | 1769.7 | 2 | R |
| 1772.7 | 1778.4 | 3 | R |
| 1782.6 | 1788.1 | 4 | R |
| 1792.2 |  |  | S |
| 1802.3 | 1805.2 | 5 | I |
| 1812.1 | 1816.4 | 6 | I |
| 1821.9 | 1829.9 | 7 | I |
| 1832.3 | 1837.2 | 8 | R |
| 1842.0 | 1848.1 | 9 | R |
| 1851.8 | 1860.1 | 10 | R |
| 1861.9 | 1870.6 | 11 | R |
| 1871.6 |  |  | S |
| 1881.5 | 1883.9 | 12 | I |
| 1891.8 | 1894.1 | 13 | I |
| 1901.6 | 1907.0 | 14 | I |
| 1911.5 | 1917.6 | 15 | R |
| 1921.5 | 1928.4 | 16 | R |
| 1931.2 | 1937.4 | 17 | R |
| 1941.0 | 1947.5 | 18 | R |
| 1951.4 | 1957.9 | 19 | R |
| 1961.2 | 1968.9 | 20 | R |
| 1971.1 | 1979.9 | 21 | R |
| 1981.1 | 1989.6 | 22 | R |
| 1990.7 | 2000.4 | 23 | R |
| 2000.7 |  | 24 | R |
| 2011.0 |  | 25 | S |
| 2020.8 |  | 26 | I |

Regular Cycle (R)
Irregular Cycle (I)
Skipped Cycle (S)

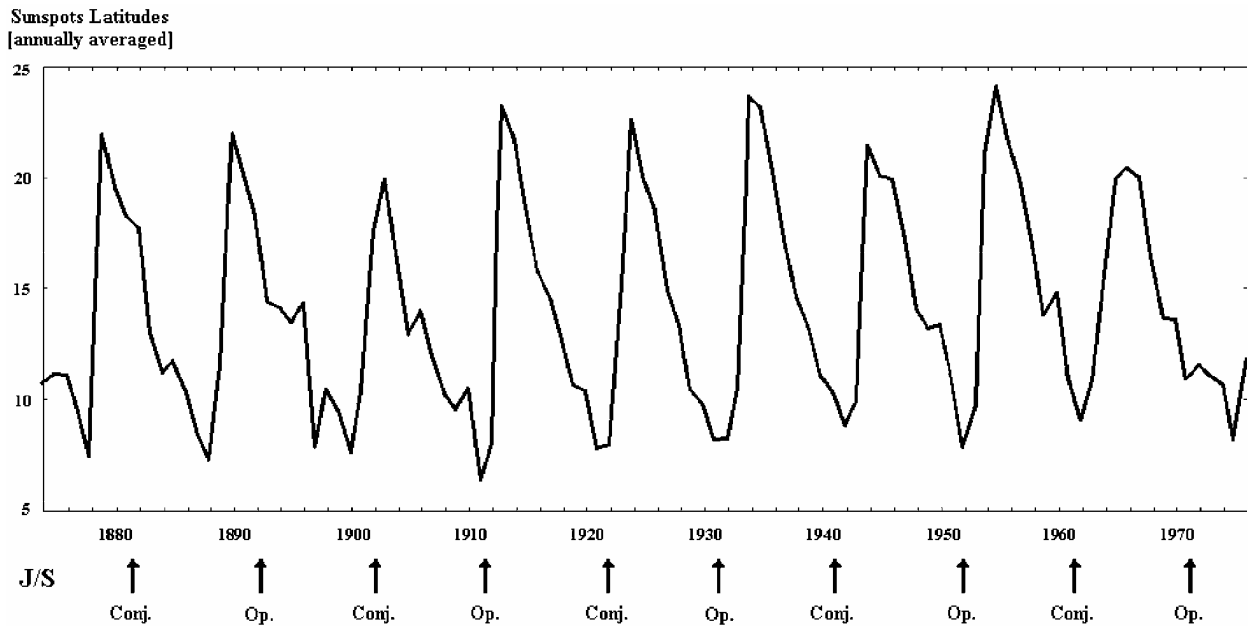

FIG. 1  Correlation between the annually averaged latitude of the sunspots (Greenwich data set from 1874 to 1976) and the heliospheric conjunctions and oppositions of Jupiter and Saturn.

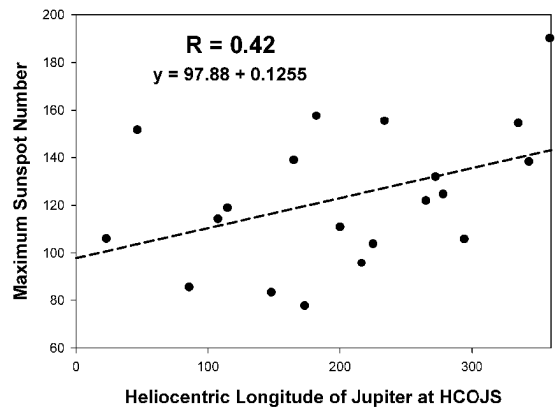

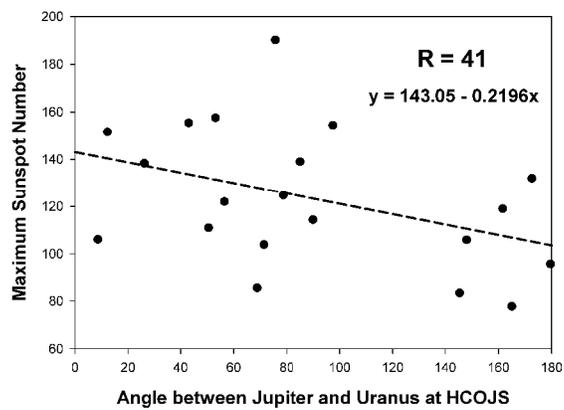

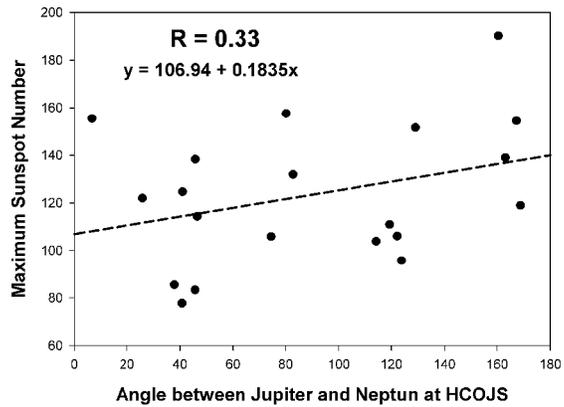

FIG. 2 Correlations between solar activity and the position of the Jovian planets at HCOJS. The solar activity is represented by the maximum mean annual sunspot number during the cycle.

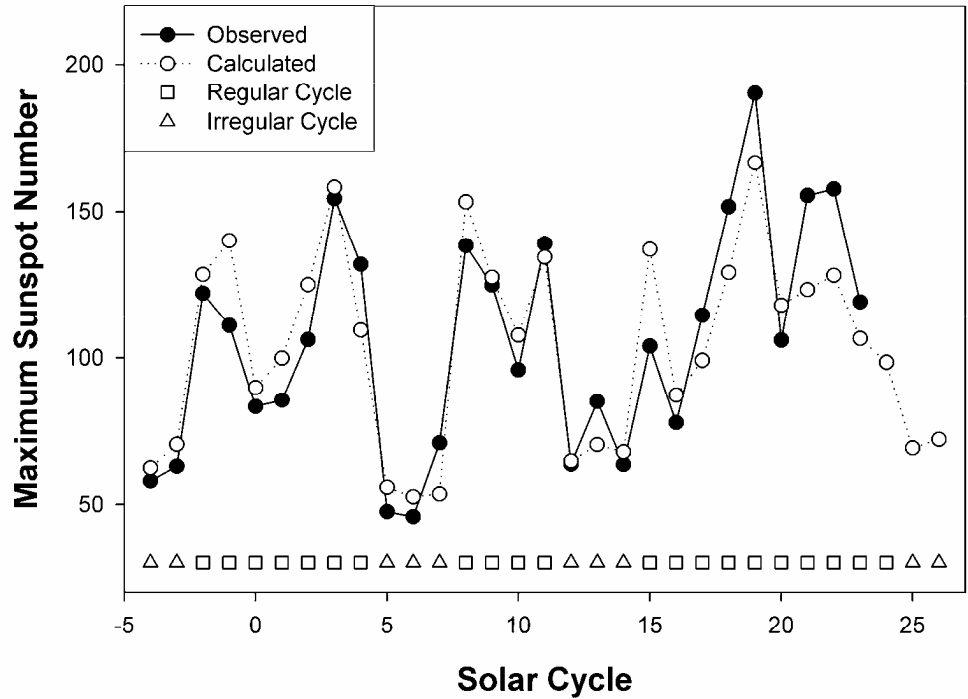

FIG. 3  Using Equations 4 and 5 the maximum mean annual sunspot number for cycles -4 through 26 are calculated.  The observed values are also plotted for comparison.